\documentclass[conference]{IEEEtran}

\newcommand{\alert}[1]{\textcolor{red}{#1}}

\long\def\comment#1{}

\newcommand{\be}{\begin{equation}}
\newcommand{\ee}{\end{equation}}

\usepackage{amsthm}
\newtheorem{theorem}{Theorem}
\newtheorem{corollary}{Corollary}[theorem]

\newtheorem{definition}{Definition}

\newfont{\bbb}{msbm10 scaled 700}

\newfont{\bb}{msbm10 scaled 1100}

\newcommand{\ev}{{\bf e}}

\newcommand{\rv}{{\bf r}}
\newcommand{\sv}{{\bf s}}

\newcommand{\xv}{{\bf x}}

\newcommand{\Am}{{\bf A}}

\newcommand{\Hm}{{\bf H}}

\newcommand{\Sm}{{\bf S}}

\newcommand{\Ac}{{\cal A}}

\newcommand{\Kc}{{\cal K}}

\newcommand{\Nc}{{\cal N}}

\newcommand{\Pc}{{\cal P}}

\newcommand{\Sc}{{\cal S}}

\newcommand{\RNum}[1]{\uppercase\expandafter{\romannumeral #1\relax}}

\newcommand{\Sigmam}{\hbox{\boldmath$\Sigma$}}

\newcommand{\eqdef}{\stackrel{\Delta}{=}}

 \usepackage{algorithm}
 \usepackage{algpseudocode}

 \usepackage{graphicx}
\usepackage{epstopdf}
 \usepackage{amssymb}
 \usepackage[english]{babel}
 \usepackage{amsmath}
 \usepackage{dsfont}
\newtheorem{prop}{Proposition}
\DeclareMathOperator*{\argmin}{arg\,min}

\usepackage[usenames, dvipsnames]{color}
\usepackage{mhchem} 
\IEEEoverridecommandlockouts
\begin{document}

\title{Power Injection Measurements are more Vulnerable to Data Integrity Attacks than Power Flow Measurements \\
\thanks{This research was supported in part by the European Commission through the H2020-MSCA-RISE-2019 program under grant 872172 and in part by the China Scholarship Council.}
}

\author{\IEEEauthorblockN{Xiuzhen Ye$^*$, I\~naki Esnaola$^{*\dag}$, Samir M. Perlaza $^{\S\dag}$, and Robert F. Harrison$^{*}$}

\IEEEauthorblockA{$^*$Dept. of Automatic  Control and Systems Engineering, University of Sheffield, Sheffield S1 3JD, UK\\
 $^\dag$Dept. of Electrical Engineering, Princeton University, Princeton, NJ 08544, USA\\
 $^\S$INRIA, Sophia Antipolis 06902, France\\
}}

\maketitle
\begin{abstract}
 A novel metric that describes the vulnerability of the measurements in power system to data integrity attacks is proposed. The new metric, coined vulnerability index (VuIx), leverages information theoretic measures to assess the attack effect on the fundamental limits of the disruption and detection tradeoff. The result of computing the VuIx of the measurements in the system yields an ordering of the measurements vulnerability based on the level of exposure to data integrity attacks. This new framework is used to assess the measurements vulnerability of IEEE test systems and it is observed that power injection measurements are overwhelmingly more vulnerable to data integrity attacks than power flow measurements. A detailed numerical evaluation of the VuIx values for IEEE test systems is provided.
\end{abstract}
 
\section{Introduction}
Supervisory Control and Data Acquisition (SCADA) systems and more recently advanced communication systems facilitate efficient, economic and reliable operation of power systems~\cite{GJ_PSanalysis_1994}. For instance, the communication system transmits the measurements to a state estimator that evaluates the operation status of the system accurately~\cite{AA_PSstateestimation_04}. However, the integration between the physical layer and the cyber layer exposes the system to cybersecurity threats. 
Cyber incidents highlight the vulnerability of power systems to sophisticated attacks that impinge multiple security definitions. To ensure the security and reliability of power system operation, it is essential to quantitatively characterize the vulnerabilities of the system in order to set up appropriate security mechanisms~\cite{WZ_CN_13}. To that end, security metrics provide operationally meaningful vulnerability descriptors and identify the impact that security threats pose on the system. Moreover, security metrics enable operators to assess the defence mechanisms requirements to be embedded into cybersecurity policies, processes, and technology~\cite{JA_PearsonEducation_07}. For example, the Common Vulnerability Scoring System (CVSS) analysis Information Technology (IT) system~\cite{M_CVSS_11}. Typical security metrics for power systems focus on integrity, availability, and confidentiality as envisioned by the cybersecurity working group in the NIST Smart Grid interoperability panel~\cite{NISTIR_securityguideline_10}. 
System security objectives are categorized into system vulnerability, defence power, attack severity, and situations to develop security metrics in a systematic manner~\cite{PM_ACM_16}.
A cyberphysical security assessment metric (CP-SAM) based on quantitative factors is proposed to assess the specific security challenges of microgrid systems~\cite{VV_TSG_19}.

This fragmented landscape showcases a wide variety of metrics available that depend on the security services, threat characteristics, and system parameters. Remarkably, there is a lack of general data integrity vulnerability metrics for power systems. For instance, the impact of data injection attacks (DIAs)~\cite{LY_TISSEC_11} can be assessed with a wide variety of criteria that depends on the objective of the attacker~\cite{CKKPT_SPM_12},~\cite{MO_JSAC_13},~\cite{IE_TSG_16}. 


 In this paper, we propose a fundamental metric to assess the vulnerability of measurements in power systems to data integrity attacks. Our aim is to provide a metric that is grounded on fundamental principles, and therefore, inform the threats and vulnerabilities of a wide range of services and processes. Specifically, we adopt an information theoretic framework to characterize the fundamental information loss induced by data integrity attacks~\cite{SE_TSG_19},~\cite{YE_SGC_20}.


A brief description of notation follows. Consider a matrix $\Am \in \mathds{R}^{m \times n}$, then $(\Am)_{ij}$ denotes the entry in row $i$ and column $j$ and $\textnormal{diag}(\Am)$ denotes the vector formed by the diagonal entries of $\Am$. The elementary vector $\ev_i$ is a vector of zeros with a one in position $i$. The set of positive semidefinite matrices of size $n \times n$ is denoted by $\Sc_+^n$.

\section{System model}\label{system model}
\subsection{Observation Model}
In a power system the state vector $\xv \in{\mathds{R}^n}$ that contains the voltages and angles at all the generations and load buses describes the operation state of the system. State vector $\xv$ is observed by the acquisition function $F: {\mathds{R}^n} \rightarrow {\mathds{R}^m}$. A linearized observation model is considered for state estimation, which yields the observation model as 
\begin{equation}\label{eq:obs_noattack}
  Y^m  = \Hm\xv+Z^m,
\end{equation}
where $\Hm \in {\mathds{R}^{m \times n}}$ is the Jacobian of the function $F$ at a given operating point and is determined by the parameters and topology of the system. The vector of measurements $Y^m$ is corrupted by additive white Gaussian noise introduced by the sensors~\cite{GJ_PSanalysis_1994},~\cite{AA_PSstateestimation_04}. The noise vector $Z^m$ follows a multivariate Gaussian distribution, that is,
\be\label{Z_distribution}
Z^m \sim \mathcal{N}(\textbf{0},\sigma^2 \textrm{\textbf{I}}_m),
\ee
where $\sigma^2$ is the noise variance.

In a Bayesian estimation framework, the state variables are described by a vector of random variables $X^n$ with a given distribution. In this study, we assume $X^n$ follows a multivariable Gaussian distribution~\cite{GE_TSG_18} with zero mean and covariance matrix $\Sigmam_{X\!X} \in S_{+}^n$, that is,
\be\label{Sigma_xx}
X^n \sim {\cal N}(\textbf{0},\Sigmam_{X\!X}).
\ee
From~(\ref{eq:obs_noattack}), it follows that the vector of measurements is with zero mean and covariance matrix $\Sigmam_{Y\!Y} \in S_{+}^m$, that is, 
\be\label{Y_Sigma_YY}
Y^m \sim {\cal N}(\textbf{0},\Sigmam_{Y\!Y}),
\ee
\vspace{-2mm}
where 
\be\label{20220706_1}
\Sigmam_{Y\!Y} \eqdef \Hm\Sigmam_{X\!X}\Hm^{\sf{T}}+\sigma^2 \textrm{\textbf{I}}_m.
\ee
\subsection{Attack Setting}
Let us denote the measurements corrupted by the malicious attack vector $A^m \in \mathds{R}^m$ as $Y_A^m$, that is,
\begin{equation}\label{obs_attack}
	Y_A^m  = \Hm X^n+Z^m + A^m,
\end{equation}
where $A^m \sim P_{A^m}$ is the distribution of the random attack vector $A^m$.
Since the Gaussian distribution minimizes the mutual information between the state variables and the compromised measurements {with} a fixed covariance matrix~\cite{SI_TIT_13}, we adopt a Gaussian random attack {framework} given by
\begin{equation}\label{eq:Gauss_attack}
  A^m \sim \mathcal{N} (\textbf{0}, \Sigmam_{A\!A}),
\end{equation}
where $\Sigmam_{A\!A} \in \Sc_+^m$ is the covariance matrix of attack vector $A^m$. Consequently, the vector of compromised measurements $Y_A^m$ follows a multivariate Gaussian distribution with zero mean and covariance matrix $\Sigmam_{Y_A\!Y_A} \in \Sc_+^m$, that is,
\begin{equation}\label{eq:obs_attack} 
  Y^m_A  \sim \mathcal{N} (\textbf{0},\Sigmam_{Y_A\!Y_A}),
\end{equation}
with $\Sigmam_{Y_A\!Y_A} \eqdef \Hm\Sigmam_{X\!X}\Hm^{\sf{T}} + \sigma^2 \textrm{\textbf{I}}_m + \Sigmam_{A\!A}$.
 
\section{Information Theoretic Attacks}\label{sparse construction}
The aim of the attack is two fold. Firstly, the attack aims to disrupt the state
estimation procedure or in general, any procedures that require measurements from the system. Secondly, it aims to stay undetected. For the first objective, instead of assuming any particular state estimation methods, we minimize the mutual information between the vector of state variables $X^n$ in~\eqref{Sigma_xx} and the vector of compromised measurements $Y_A^m$ in~\eqref{obs_attack}, that is, $I(X^n;Y_A^m)$. 
In the other words, the attack yields less information about the state variables contained by the compromised measurements.
The stealthy constraint in the second objective is captured by the Kullback Leibler (KL) divergence between the distribution $P_{Y^m_A}$ in~\eqref{obs_attack} and the distribution $P_{Y^m}$ in~\eqref{eq:obs_noattack}, that is, $D(P_{Y_A^m}\|P_{Y^m})$. The Chernoff-Stein lemma~\cite{book_EIT} states that the minimization of KL divergence leads to the minimization of the asymptotic detection probability. 

The following propositions characterize mutual information and KL divergence in Gaussian state variables and attacks, respectively.

\begin{prop}\cite[Prop. 2]{SE_TSG_19}\label{Prop_MI}
	The mutual information between the vectors of random variables $X^n$ in~\eqref{Sigma_xx} and $Y_A^m$ in~\eqref{obs_attack} is 
	\begin{equation} 
		\begin{aligned}\label{eq_MI}
			I(X^n;Y^m_A) 
			= &\frac{1}{2}\textnormal{log}  \left|\Hm \Sigmam_{X\!X}\Hm^{\sf T}\left(\Sigmam_{A\!A} + \sigma^2\textbf{I}_m \right)^{-1}     + \textbf{I}_m \right|,
		\end{aligned}
	\end{equation}	
	where the matrix $\Hm $ is in~\eqref{eq:obs_noattack}, the real $\sigma \in \mathds{R}_+$ is in~\eqref{Z_distribution}, the matrix $\Sigmam_{X\!X} $ is in~\eqref{Sigma_xx} and the matrix $\Sigmam_{A\!A}$ is in \eqref{eq:Gauss_attack}.
\end{prop}

\begin{prop}\cite[Prop. 1]{SE_TSG_19}\label{Prop_KL}
	The KL divergence between the distribution of random vector $Y_A^m$ in~\eqref{obs_attack} and the distribution of random vector $Y^m$ in~\eqref{Y_Sigma_YY} is 
	\begin{equation} 
		\begin{aligned}\label{eq_KL} 
			D(P_{Y_A^m}\|P_{Y^m}) 
			= & \frac{1}{2}\left(-\textnormal{log}  \left|  \textbf{I}_m + \Sigmam_{Y\!Y}^{-1}\Sigmam_{A\!A}\right|  + \textnormal{tr}\left(\Sigmam_{Y\!Y}^{-1}\Sigmam_{A\!A} \right) \right),
		\end{aligned}
	\end{equation}
	where the matrices $\Sigmam_{Y\!Y}$ and $\Sigmam_{A\!A}$ are in~\eqref{20220706_1} and \eqref{eq:Gauss_attack}, respectively.
\end{prop}

Particularly, the attack construction is proposed in the following optimization problem~\cite{SE_TSG_19}~\cite{YE_SGC_20}:
\begin{equation}\label{eq:stealth_opt}
	\min_{P_{A^m}} I(X^n;Y^m_A)+ \lambda D(P_{Y_A^m}\|P_{Y^m}),
\end{equation}
where $\lambda \in \mathds{R}_+$ is the weighting parameter that determines the tradeoff between mutual information and KL divergence. 
Note that the optimization domain in~\eqref{eq:stealth_opt} is the set of $m$-dimensional Gaussian multivariate distributions. The optimal Gaussian attack for $\lambda \leq 1$ as a solution to~\eqref{eq:stealth_opt} is given by~\cite{SE_TSG_19} 
\be\label{opt_stealth}
P_{A^m}^* \sim\Nc(\mathbf{0},{\lambda^{-1/2}}\Hm\Sigmam_{X\!X}\Hm^{\sf T}).
\ee
Note that the attack realizations from~\eqref{opt_stealth} are nonzero with probability 1, that is, $\mathds{P}[\left| \textnormal{supp}({A^m})\right |= m] = 1$, where
\be
\textnormal{supp}({A^m})\eqdef\left\{i:\mathds{P}\left[A_i=0\right]=0\right\}.
\ee
In~\cite{YE_SGC_20}, the attack construction incorporates the sparsity constraint by limiting the optimization domain over the attack vector $A^m$ in~\eqref{obs_attack} to the distributions with cardinality of the support satisfying $|\textnormal{supp}({A^m})|=k\leq m$, that is,
\vspace{-1.5mm}
\be
\label{EqSparcityDomain}
\Pc_k\eqdef\bigcup_{i=1}^k\left \{{A^m}\sim \Nc(\mathbf{0},\bar{\Sigmam}):\left| \textnormal{supp}({A^m})\right | =i\right \}.
\ee
\vspace{-2mm}
The resulting attack construction with the sparsity constraints is
\begin{equation}\label{eq:sparsestealth_opt} 
	\min_{\Pc_k} I(X^n;Y^m_A)+ \lambda D(P_{Y_A^m}\|P_{Y^m}).
\end{equation}
The following theorem provides the optimal single sensor attack.

\begin{theorem}\cite[Th. 1]{YE_SGC_20}\label{Theorem_singlesensor} 
	The solution to the sparse stealth attack construction problem in~\eqref{eq:sparsestealth_opt}  for the case $k=1$ is 
	\begin{equation}\label{26}
		\bar{\Sigmam}^*=v\ev_{i}\ev_{i}^{\sf T},
	\end{equation}
\vspace{-3mm}
	where
	\begin{subequations} 
		\begin{align}
			i&= \argmin_{j \in \{1,2,\ldots,m\}} \left\{\left(\Sigmam_{Y\!Y}^{-1}\right)_{jj}\right\} \label{eq:opt_i}\\
			v&=- \frac{\sigma^2}{2} + \frac{1}{2}\left( \sigma^4 -\frac{4 (\underbar{$w$}\sigma^2 -1)}{\lambda \underbar{$w$}^2} \right)^{\frac{1}{2}} \label{eq:opt_v}
		\end{align}
	\end{subequations}
	with $\underbar{$w$}\eqdef (\Sigmam_{Y\!Y}^{-1})_{ii}$.
\end{theorem}

 \vspace{-3mm}
\section{Vulnerability of Measurements to Information Theoretic Attacks}
\subsection{Attack Structure with Sequential Sensor Selection}
To assess the impact of the attacks to different measurements, we model the entries of the attack vector with independency, that is, 
\begin{equation}
	\label{EqIndepPA}
	P_{A^m}=\prod_{i=1}^m P_{A_i},
\end{equation} 
where for all $i \in \lbrace 1,2, \ldots, m\rbrace$, the distribution $P_{A_i}$ is Gaussian with zero mean and variance $v \in \mathds{R}_+$. 
Consider that $k$ sensors have been compromised with $k \in \{1,2,\ldots,m-1\}$ and let the covariance matrix of the corresponding attack vector $A^m$ in~\eqref{obs_attack} be
\be\label{20220310_1}
\Sigmam \in \Sc_k,
\ee
where $\Sc_k$ is the set of $m$-dimensional positive semidefinite matrix with $k$ nonzero entries in the diagonal, that is,
\be\label{20220609_1}
\Sc_k \eqdef \{\Sm \in \Sc_+^m:\|\textnormal{diag}(\Sm)\|_0 = k\}.
\ee
Let the set of sensors that have not been compromised be
	\begin{align}\label{set_K_o}
{\cal K}_o = & \lbrace i \in \lbrace 1,2, \ldots, m \rbrace: (\Sigmam)_{ii} = 0 \rbrace.
	\end{align}
The sequential sensor selection imposes the following structure in the covariance matrix of the attack vector:
\be\label{20220312_1}
\Sigmam_{A\!A} = \Sigmam+v\ev_i \ev_i^{\sf T},
\ee
where $i\in\Kc_o$ and $v\in\mathds{R}_+$.
The cost function $f: \Sc_k \times \mathds{R}_+ \times \mathds{R}_+ \times {\cal K}_o \rightarrow \mathds{R}_+$ defined by adding \eqref{eq_MI} and~\eqref{eq_KL} is as follows:
\begin{align}\label{MI_f_k}
	f(\Sigmam, \lambda,v, i) \alert{\eqdef} & I(X^n;Y^m_A) +\lambda D(P_{Y_A^m}\|P_{Y^m}) \\
	= & \frac{1}{2}(1-\lambda)\textnormal{log}  \left| \Sigmam_{Y\!Y} + \Sigmam + v\ev_i \ev_i^{\sf T} \right| \\
		\nonumber
		&- \frac{1}{2}\textnormal{log}  \left|\Sigmam + v\ev_i \ev_i^{\sf T} +\sigma^2 \textbf{I}_m\right| \\
		\nonumber
	&+\frac{1}{2} \lambda \left( \textnormal{tr}\left(\Sigmam_{Y\!Y}^{-1}\left(\Sigmam + v\ev_i \ev_i^{\sf T}\right) \right) + \log \left| \Sigmam_{Y\!Y}\right|\right). 
\end{align}

\subsection{Information theoretic vulnerability of a measurement}

We propose a notion of vulnerability that is linked to the information theoretic cost function proposed in \cite{SE_TSG_19} to characterize the disruption and detection tradeoff incurred by the attacks. Taking the state of the system with $k$ compromised measurements as the baseline, we quantify the vulnerability of each measurement in terms of the cost decrease induced by attacking a sensor $i$ with $i\in {\cal K}_o$. 
In the following, we define the vulnerability of a measurement.
\begin{definition}
\label{def:vul_unc}
The function $\Delta: \Sc^m_+ \times \mathds{R}_+ \times \mathds{R}_+ \times \Kc_o \rightarrow \mathds{R}_+$, where $\Kc_o$ is in~\eqref{set_K_o}, defines the {\it vulnerability of measurement $i$} in the following form:
 	\be
	\label{def_Delta_i}
 	\Delta(\Sigmam, \lambda,v, i) \eqdef f(\Sigmam, \lambda,v, i) -f(\Sigmam, \lambda,0, i),
 	\ee
 where the function $f$ is defined in~\eqref{MI_f_k}.
\end{definition}
Note that the attacker aims to minimize~\eqref{MI_f_k} by choosing an index $i$ and a variance $v$, and therefore, the definition above implies that given that $k$ sensors in $\{1,2,\ldots,m\}\setminus\Kc_o$ are already attacked in the system, the most vulnerable measurement is obtained by solving the following optimization problem
\be\label{20020310_13}
\min_{i \in {\cal K}_o} \Delta(\Sigmam, \lambda, v, i),
\ee
where $\Kc_o$ is defined in~\eqref{set_K_o}.
\subsection{Vulnerability analysis of uncompromised systems}\label{Vulnerability analysis on single sensor attack}
 We first consider the case in which no sensors are under attack, that is, $k=0$, and the attacker selects a single sensor and corrupts the corresponding measurement with a given budget $v\leq v_0$. We quantify the vulnerability of measurement $i$ in terms of $\Delta(\Sigmam, \lambda, v, i)$.
%
%

For the uncompromised system case, the optimization problem in \eqref{20020310_13} can be solved in closed form expression. The following theorem provides the solution.
\begin{theorem}\label{Theorem_vulnerability_singlesensor}
 The solution to the problem in~\eqref{20020310_13}, with $\Kc_o = \{1,2,\ldots,m\}$, is
		\be\label{20220603_2}
			i= \argmin_{j \in \{1,2,\ldots,m\}} \left\{\left(\Sigmam_{Y\!Y}^{-1}\right)_{jj}\right\},
	    \ee
	    where $\Sigmam_{Y\!Y}$ is in~\eqref{20220706_1}.
\end{theorem}
\begin{proof}
	We start by noting that $f (\textbf{0}, \lambda, 0, i)$ is a constant with respect to $i$. Hence, optimization problem in~\eqref{20020310_13} is equivalent to
	\be\label{20220310_13}
	\min_{i \in \{1,2,\ldots,m\}} f(\textbf{0}, \lambda,v, i).
	\ee
	Let $\lambda \in \mathds{R}_+$ and $v \in \mathds{R}_+$. The resulting problem in~\eqref{20220310_13} is equivalent to the following optimization problem:
	\begin{align}\label{20220603_1} 
		\min_{i \in \{1,2,\ldots,m\}} & (1\!-\!\lambda)\textnormal{log}  \left( 1 \!+ \!v \textnormal{tr}\left(\Sigmam_{Y\!Y}^{-1}\ev_i \ev_i^{\sf T} \right)\right) \!+ \!\lambda v \textnormal{tr}\left(\Sigmam_{Y\!Y}^{-1}\ev_i \ev_i^{\sf T} \right)\!.
	\end{align}
The proof concludes by noting that the cost function in~\eqref{20220603_1} is monotonically increasing with respect to $\textnormal{tr}\left(\Sigmam_{Y\!Y}^{-1}\ev_i \ev_i^{\sf T} \right)$.
\end{proof}
From Theorem~\ref{Theorem_vulnerability_singlesensor}, it follows that the identification of the most vulnerable measurement is independent of $\lambda$ in~\eqref{MI_f_k} and the variance $v$. That is, it exclusively depends on the system topology denoted by $\Sigmam_{Y\!Y}$ in~\eqref{20220706_1}. This result coincides with Theorem~\ref{Theorem_singlesensor} in the sense that in the attack construction for $k=1$ in~\eqref{eq:sparsestealth_opt}, the most vulnerable measurement is in~\eqref{eq:opt_i}, which is independent of the value of $\lambda$. The following corollary formalizes this observation.

\begin{corollary}\label{coro_vulnerabilityordering_k0}
	Consider the parameters $\Sigmam = \textbf{0}$, $v \in \mathds{R}_+$ and $\lambda \in \mathds{R}_+$. Let the vulnerability ranking
		\be
		\sv \eqdef (s_1, s_2, \ldots,s_m)
		\ee 
		be such that for all $i \in \{1,2,\ldots,m\}, s_i \in \{1,2,\ldots,m\}$ and 
		\be
		\textnormal{tr}\left(\Sigmam_{Y\!Y}^{-1}\ev_{s_1} \ev_{s_1}^{\sf T} \right) \leq \textnormal{tr}\left(\Sigmam_{Y\!Y}^{-1}\ev_{s_2} \ev_{s_2}^{\sf T} \right) \leq \ldots \leq \textnormal{tr}\left(\Sigmam_{Y\!Y}^{-1}\ev_{s_m} \ev_{s_m}^{\sf T} \right).
		\ee
		For all $i\in \{1,2,\ldots,m\}$, the $i$-th most vulnerable measurement is $s_i$.
\end{corollary}
\subsection{Information theoretic vulnerability index (VuIx)} 
The vulnerability analysis of uncompromised systems in Section~\ref{Vulnerability analysis on single sensor attack} is constrained to $k = 0$. To generalize the vulnerability analysis to compromised systems when $k > 0$, in the following we propose a novel metric, that is, {\it vulnerability index}, for all $i \in {\cal K}_o$.
\begin{definition}
	\label{def:vul_com}
	For $k \in \{1,2,\ldots,m-1\}$ and $\Sc_{k}$ in~\eqref{20220609_1}, consider the parameters $\Sigmam\in\Sc_{k}$, $v \in \mathds{R}_+$, $\lambda \in \mathds{R}_+$. Consider also the set $\{(i, \Delta): i \in \Kc_o \}$, with $\Kc_o$ in~\eqref{set_K_o} and $\Delta_i \eqdef \Delta(\Sigmam, \lambda, v, i)$. Let the vulnerability ranking $\rv = (r_1, r_2, \ldots, r_{|\Kc_o|})$ be such that for all $i \in \{1,2,\ldots,|\Kc_o|\}$, $r_i \in \Kc_o$ and moreover,
		\be
		\vspace{-1mm}
	\Delta_{r_1}  \leq \Delta_{r_2} \leq \ldots \leq \Delta_{r_{|\Kc_o|}}. 
	\ee
	The vulnerability index (VuIx) of measurement $r_j \in \Kc_o$ is $j$, that is, $\textnormal{VuIx}(r_j)=j$.
\end{definition}
	\vspace{-1mm}
Note that the measurement with the smallest VuIx is the most vulnerable measurement that corresponds the solution to the optimization problem in~\eqref{20020310_13}.
The proposed VuIx for $i \in {\cal K}_o$ is obtained from Algorithm~\ref{alg:vulnerability index}.
\vspace{-3mm}
\begin{algorithm}
  \caption{Computation of Vulnerability Index (VuIx)}
\label{alg:vulnerability index}
	\begin{algorithmic}[1]
		\Require $\Hm$ in \eqref{eq:obs_noattack};\newline
		$\sigma^2$ in \eqref{Z_distribution};\newline
		$\Sigmam_{X\!X}$ in \eqref{Sigma_xx};\newline
		$\Sigmam \in \Sc_k$ in \eqref{20220310_1};\newline
		$\lambda \in \mathds{R}_+$ and 
		$v \in \mathds{R}_+$.
		\Ensure the VuIx for all $i \in {\cal K}_o$.
		\State Set $\Kc_o$ in~\eqref{set_K_o}
		\For {$i \in {\cal K}_o$}
			\State Compute $\Delta(\Sigmam,\lambda,v,i)$ in~\eqref{def_Delta_i}
		\EndFor
		\State Sort $\Delta(\Sigmam, \lambda, v, i) $ in ascending order
		\State Set $\rv = (r_1, r_2, \ldots, r_{|\Kc_o|})$  
		\State Set the VuIx of measurement $r_j \in \Kc_o$ as $j$.
	\end{algorithmic}
\end{algorithm}

\vspace{-3mm}
\section{Numerical results}\label{simulation}
In this section, we numerically evaluate the VuIx of the measurements on a direct current (DC) setting for the IEEE Test systems~\cite{UoW_ITC_99}. The voltage magnitudes are set to 1.0 per unit, that is, the measurements of the systems are active power flow between the buses that are physically connected and active power injection to all the buses. The Jacobian matrix $\Hm$ in~\eqref{eq:obs_noattack} determined by the topology of the system and the physical parameter of the branches is generated by MATPOWER~\cite{matpower}. We adopt a Toeplitz model for the covariance matrix $\Sigmam_{X\!X}$ that arises in a wide range of practical settings, such as autoregressive stationary processes. Specifically, we model the correlation between state variable $X_i$ and $X_j$ with a exponential decay parameter $\rho \in \mathds{R}_+$, which results in the entries of the matrix $(\Sigmam_{X\!X})_{ij} = \rho^{|i-j|}$ with $(i,j) \in \{1,2,\ldots, n\} \times \{1,2,\ldots, n\}$.
In this setting, the VuIx of the measurements is also a function of the correlation parameter $\rho$, the noise variance $\sigma^2$, and the Jacobian matrix $\Hm$. The noise regime in the observation model is characterized by the signal to noise ratio (SNR) defined as
\be
\vspace{-1mm}
\textrm{SNR} \eqdef 10\log_{10}\left(\frac{\textrm{tr}(\Hm\Sigmam_{X\!X}\Hm^{\sf T})}{m\sigma^{2}}\right).
\ee 

For all $\lambda \in \mathds{R}_+$ and $v \in \mathds{R}_+$, we generate a realization of $k$ attacked indices $\Kc_a\subseteq\{1, 2, \ldots, m\}$ that is uniformly sampled from the set of sets given by
\be
\vspace{-2mm}
\tilde{\Kc}=\left\{\Ac\subseteq\{1, 2, \ldots, m\}: |\Ac|=k\right\}.
\ee
We then construct a random covariance matrix describing the existing attacks on the system as
\vspace{-1mm}
\be 
\tilde{\Sigmam}=\sum_{i\in{\Kc_a}}\ev_i\ev_i^{\sf T},
\vspace{-1mm}
\ee
with $\Kc_a\in\tilde{\Kc}$.
In the numerical simulation, we obtain the vulnerability of measurement $i$ by computing
\vspace{-1mm}
\be 
 \Delta(\tilde{\Sigmam}, \lambda, 1, i),
\ee
\vspace{-2mm}
where $i \in \Kc_o$ is in~\eqref{set_K_o} and $\Delta$ is defined in~\eqref{def_Delta_i}.
\begin{figure}[htbp]
	\vspace{-12mm}
	\centering
	\includegraphics[width=6.9cm]{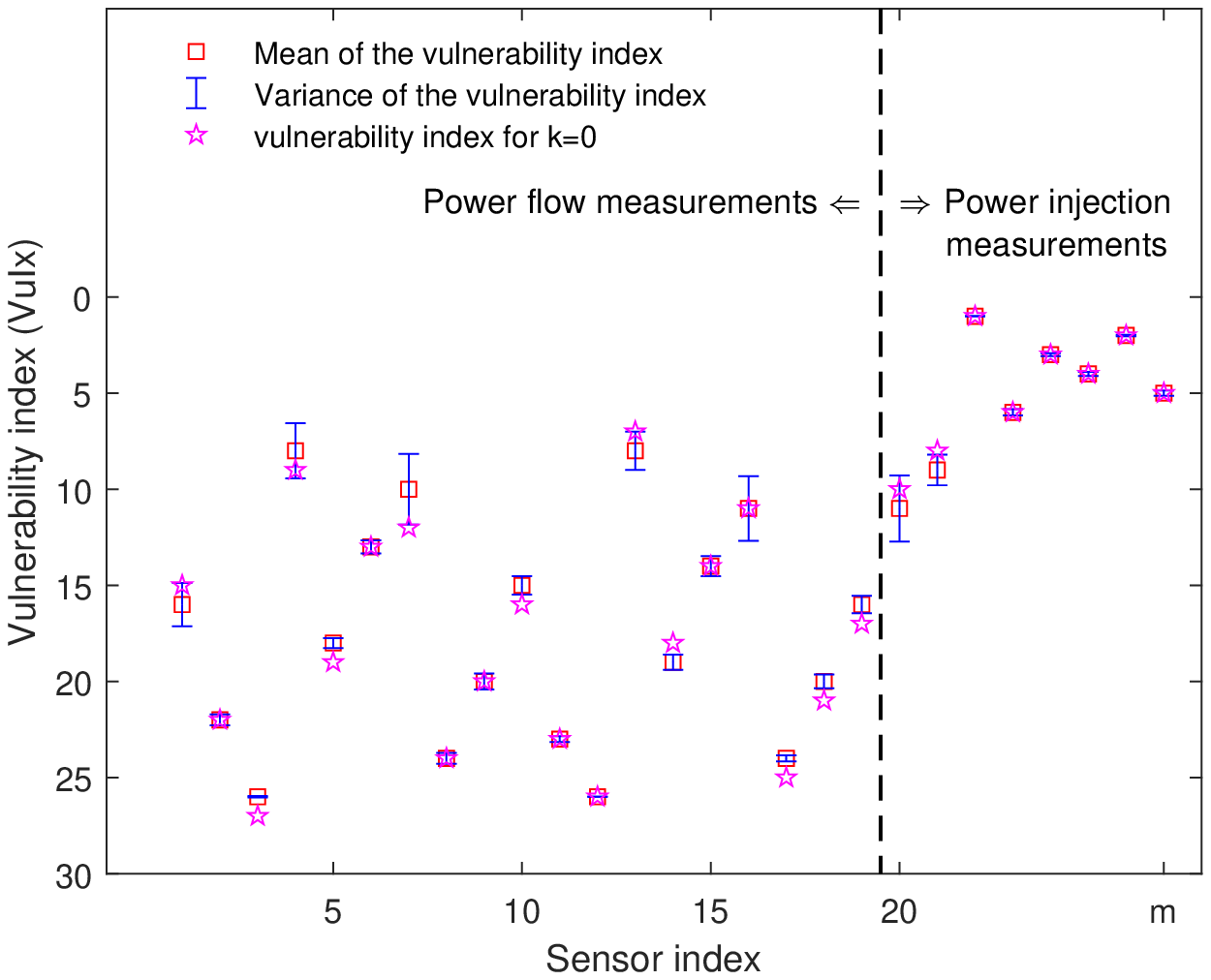}
	\caption{Vulnerability index (VuIx) when {$k = 1$, SNR = 10 dB}, $\lambda = 2$ and $\rho=0.1$ on the IEEE {9-bus system}.}\label{fig1}
	\centering
	\includegraphics[width=6.9cm]{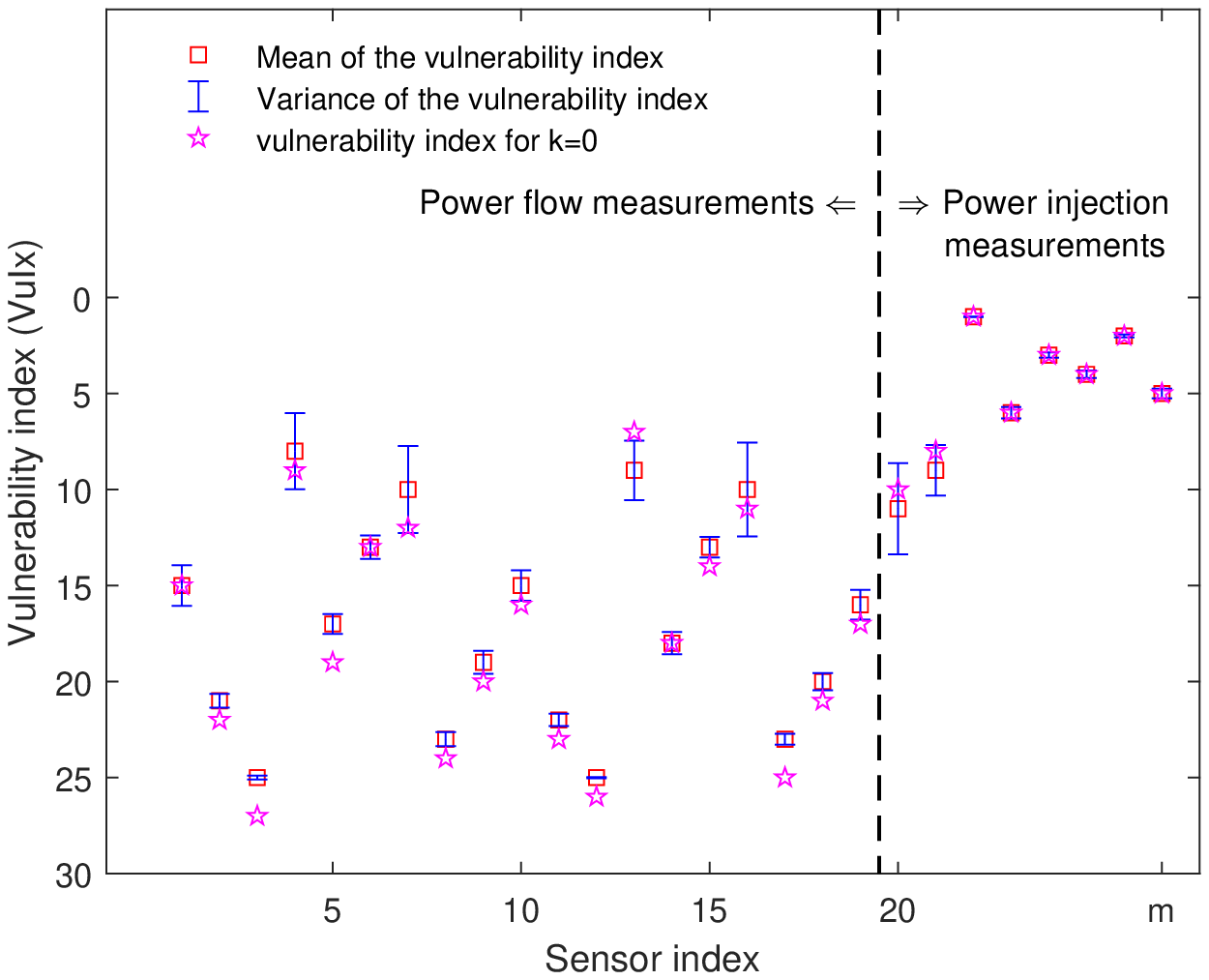}
	\caption{Vulnerability index (VuIx) when {$k = 2$, SNR = 10 dB}, $\lambda = 2$ and $\rho=0.1$ on the IEEE {9-bus system}.}\label{fig2}
	\centering
	\includegraphics[width=6.9cm]{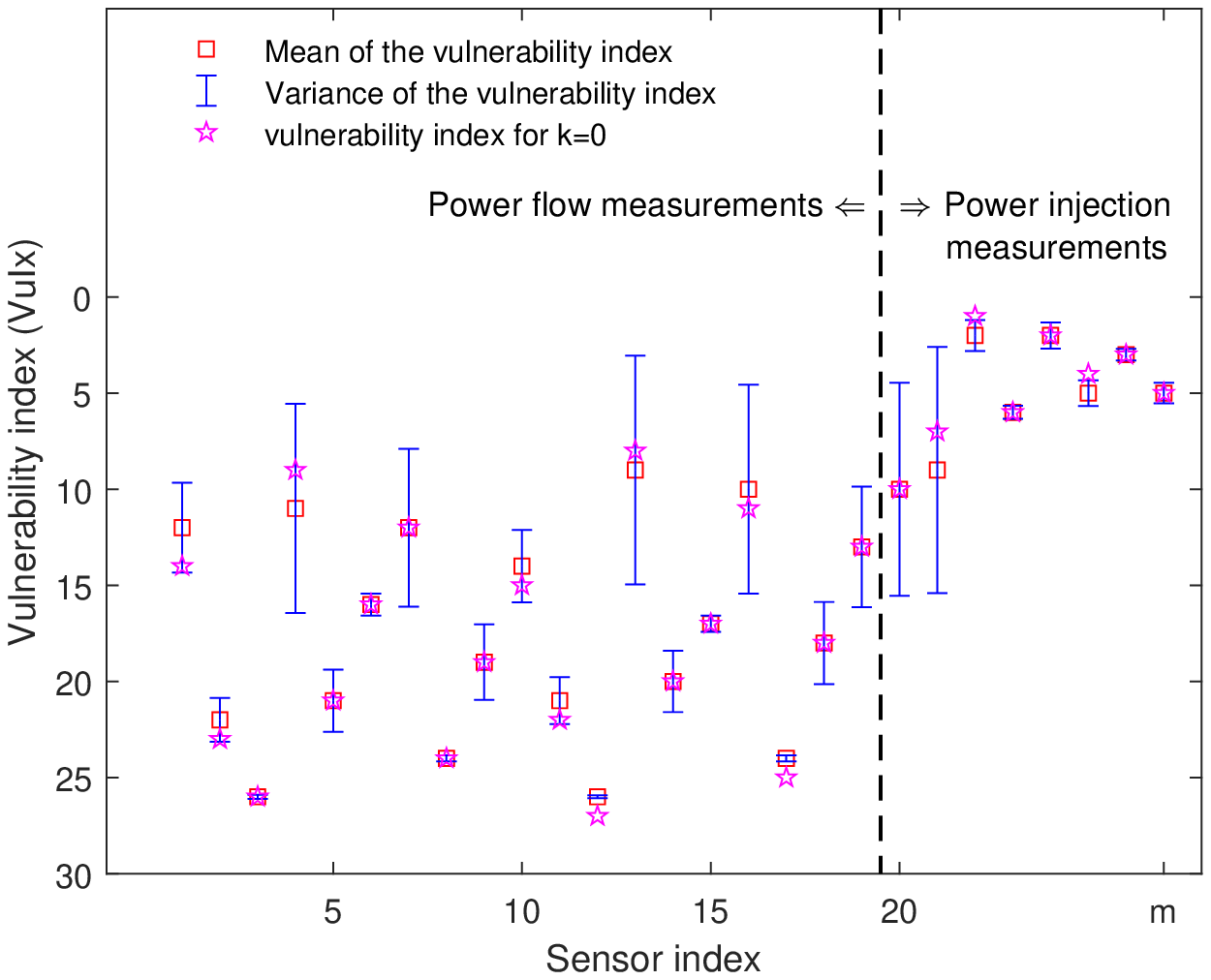}
	\caption{Vulnerability index (VuIx) when {$k = 1$, SNR = 30 dB}, $\lambda = 2$ and $\rho=0.1$ on the IEEE {9-bus system}.}\label{fig3}
	\centering
	\includegraphics[width=6.9cm]{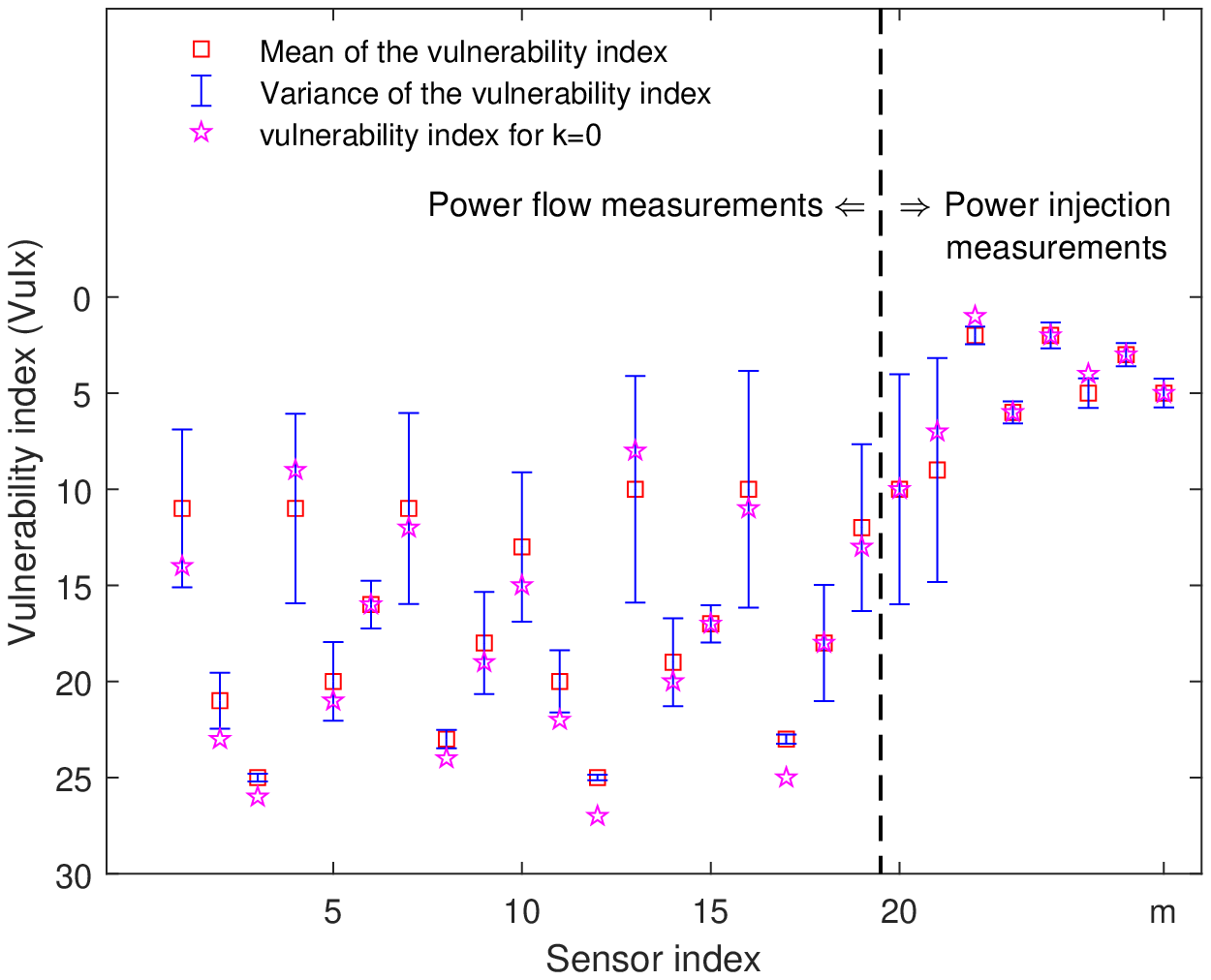}
	\caption{Vulnerability index (VuIx) when {$k = 2$, SNR = 30 dB}, $\lambda = 2$ and $\rho=0.1$ on the IEEE {9-bus system}.}\label{fig4}
\end{figure}

 \begin{figure}[htbp]
	\vspace{-12mm}
	\centering
	\includegraphics[width=6.9cm]{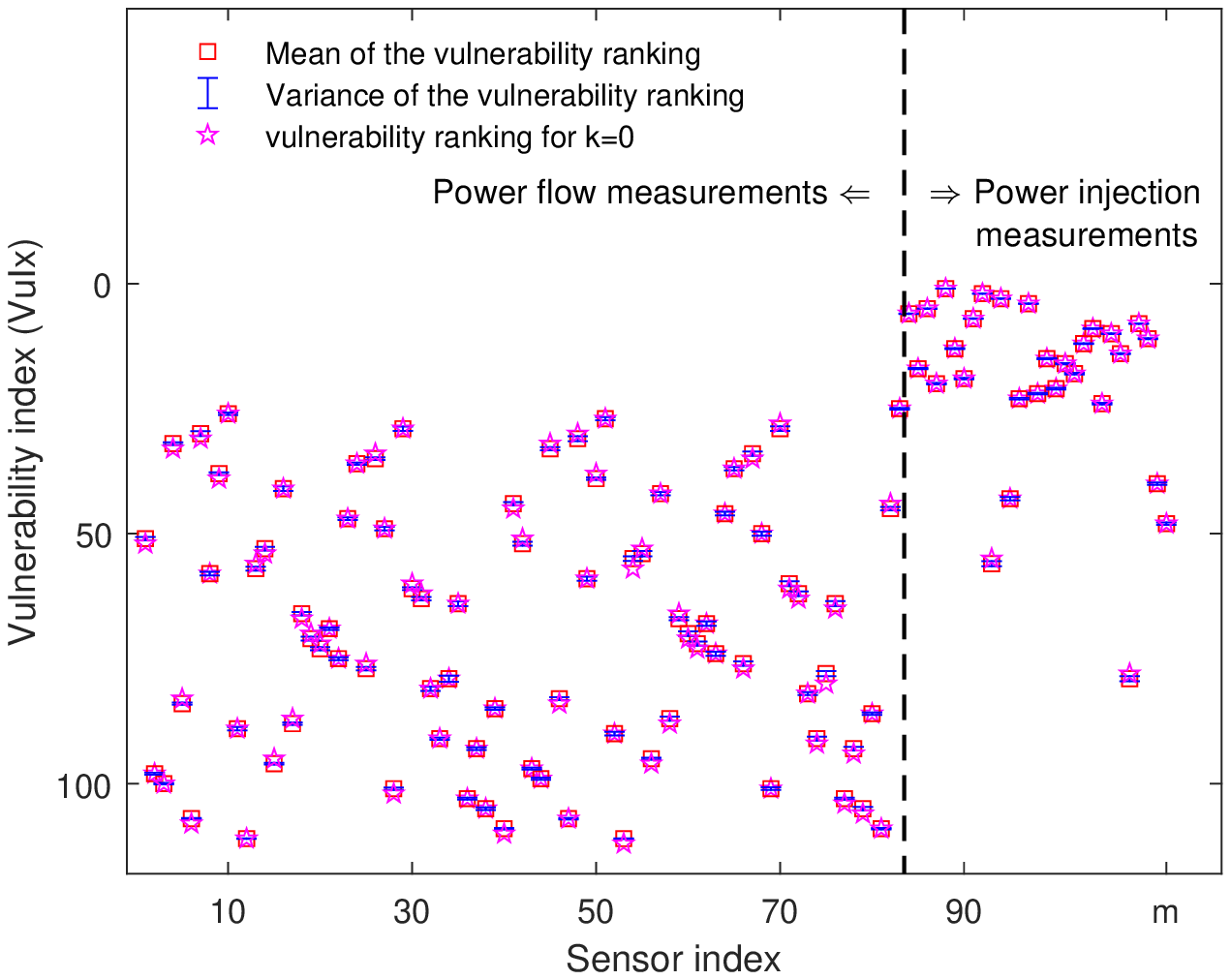}
	\caption{Vulnerability index (VuIx) when {$k = 1$}, SNR = 10 dB, $\lambda = 2$ and $\rho=0.1$ on the IEEE {30-bus system}.}\label{fig5}
	\centering
	\includegraphics[width=6.9cm]{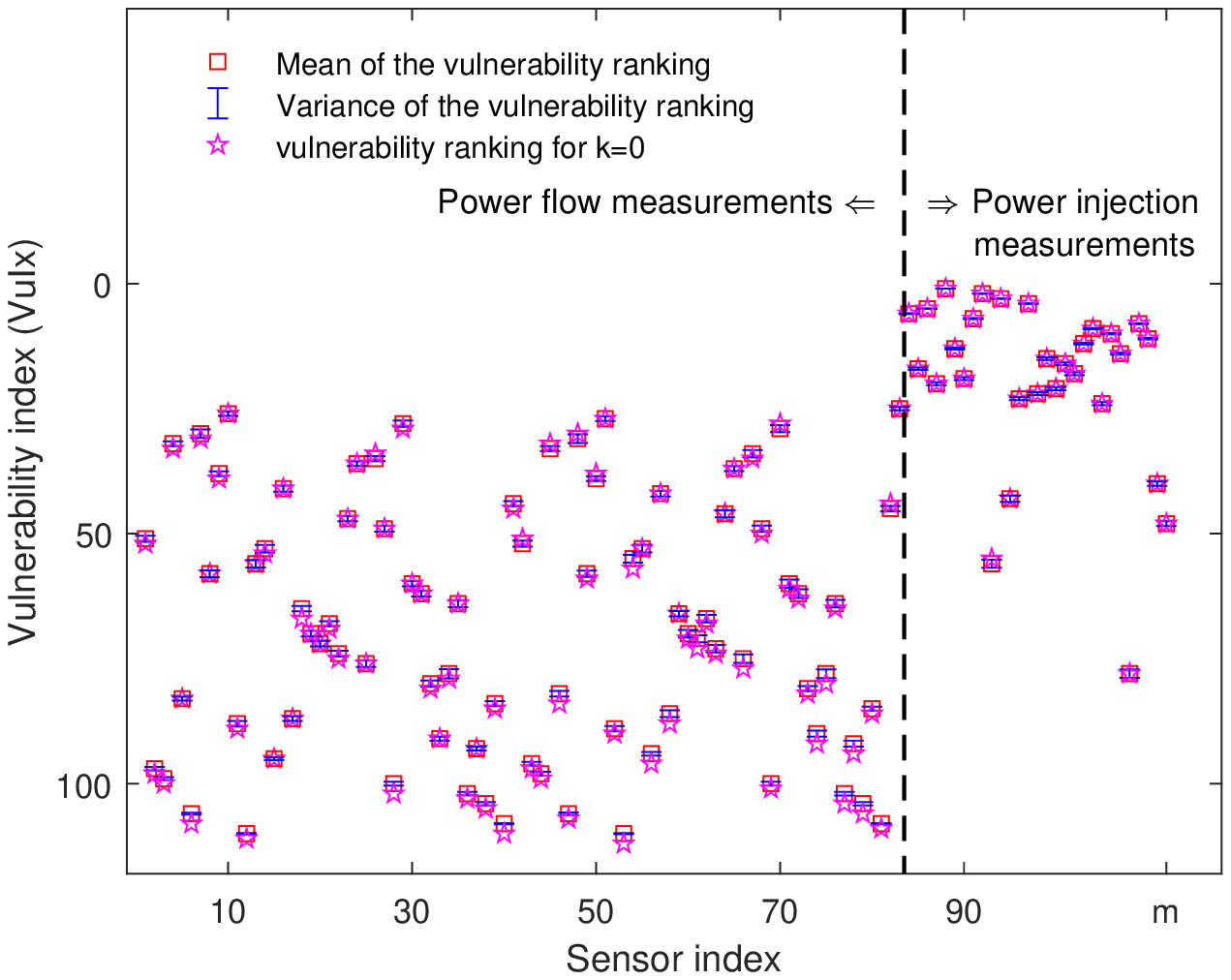}
	\caption{Vulnerability index (VuIx) when {$k = 2$}, SNR = 10 dB, $\lambda = 2$ and $\rho=0.1$ on the IEEE {30-bus system}.}\label{fig6}
	\centering
	\includegraphics[width=6.9cm]{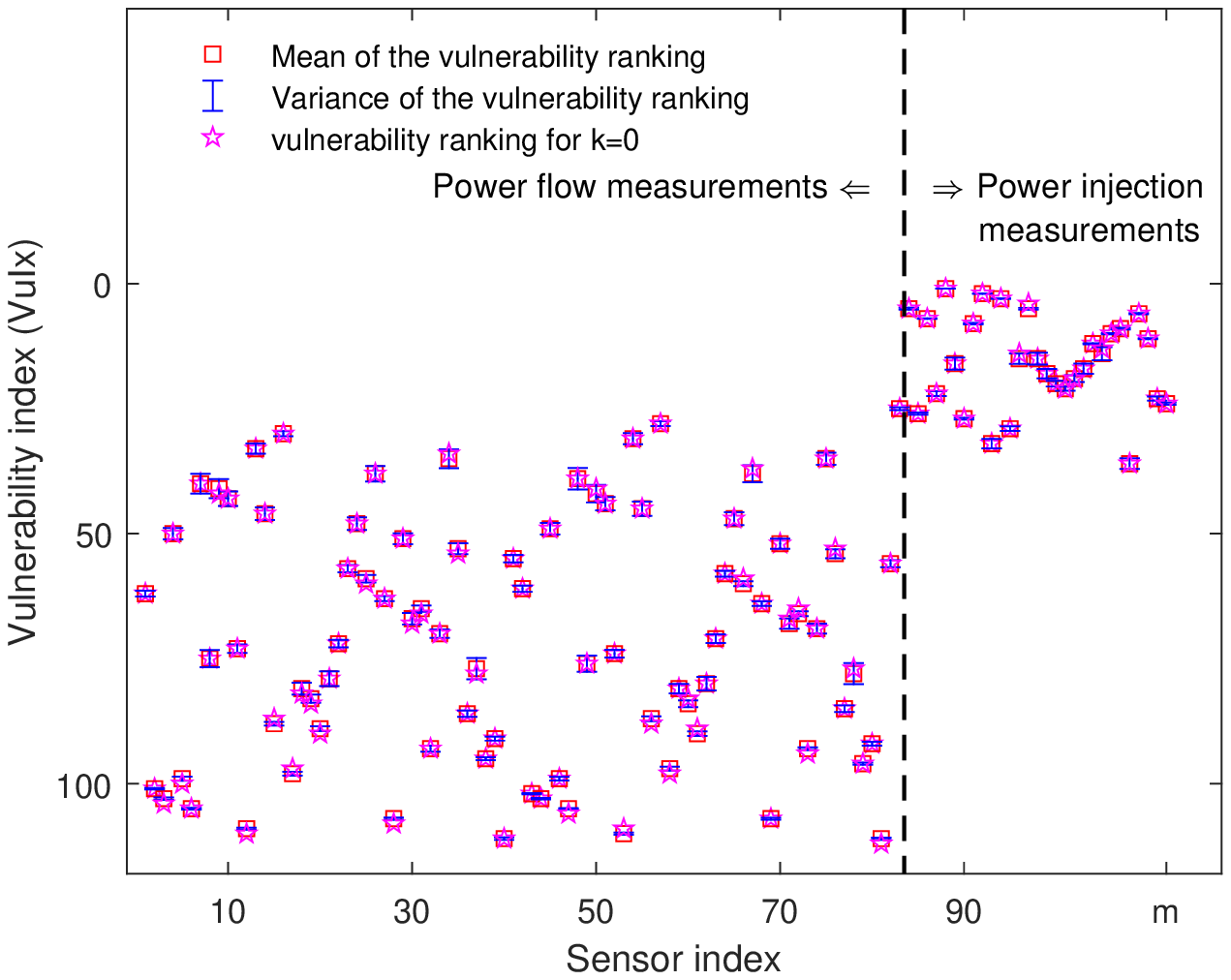}
	\caption{Vulnerability index (VuIx) when {$k = 1$}, SNR = 30 dB, $\lambda = 2$ and $\rho=0.1$ on the IEEE {30-bus system}.}\label{fig7}
	\centering
	\includegraphics[width=6.9cm]{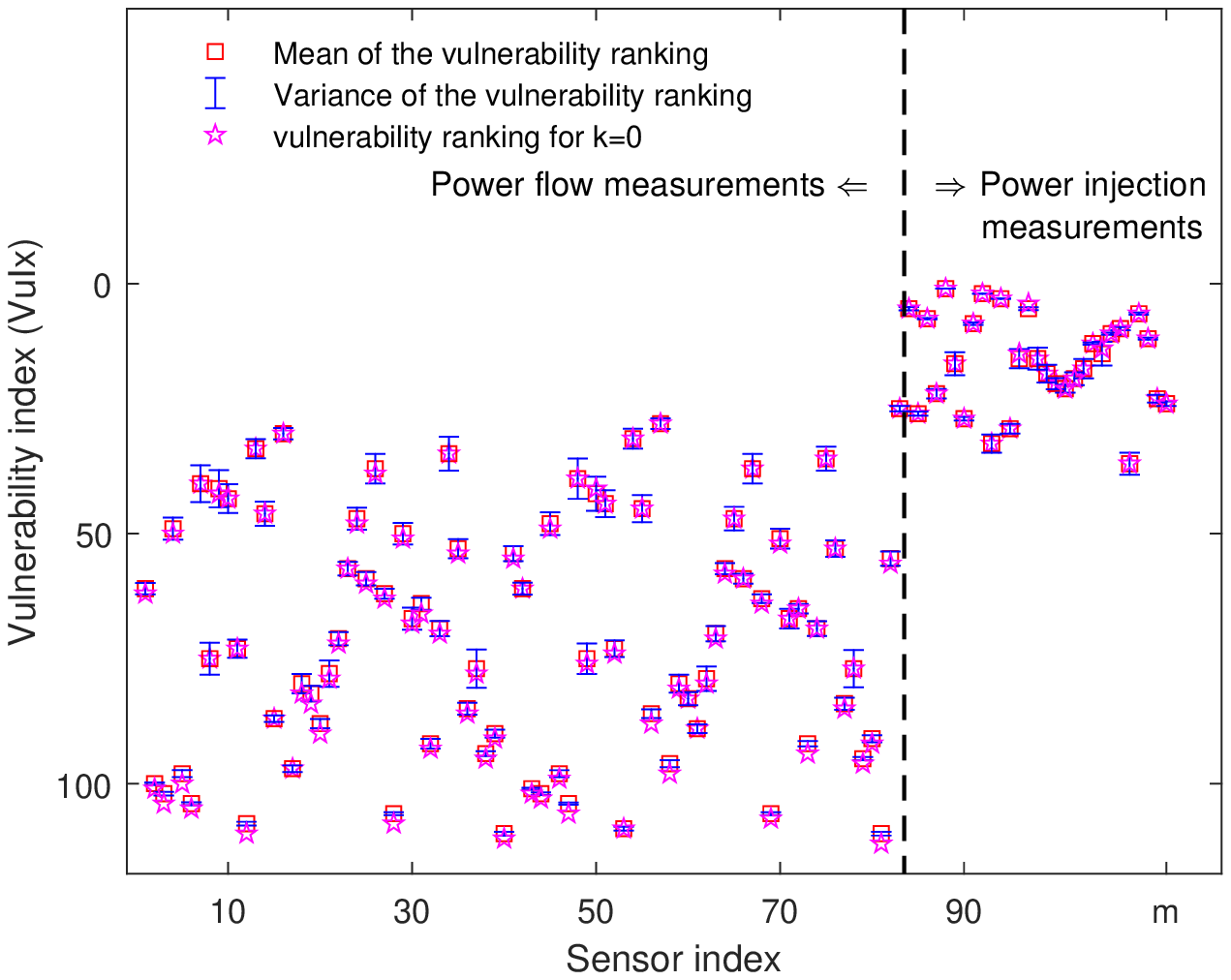}
	\caption{Vulnerability index (VuIx) when {$k = 2$}, SNR = 30 dB, $\lambda = 2$ and $\rho=0.1$ on the IEEE {30-bus system}.}\label{fig8}
\end{figure}
\begin{figure}[htbp] 
	\vspace{-5mm}
	\centering
	\includegraphics[width=6.9cm]{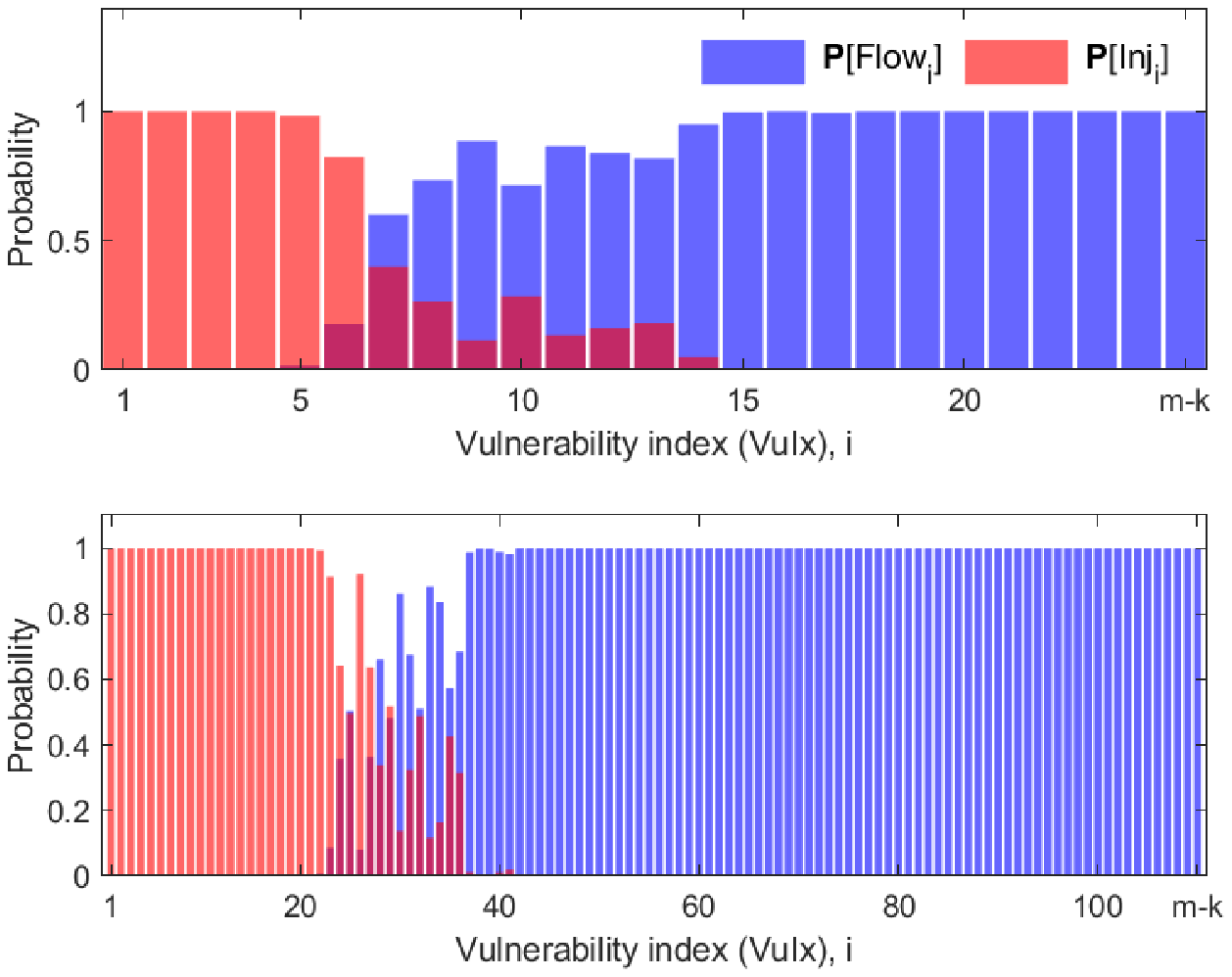}
	\caption{Probability of Vulnerability index (VuIx) corresponds to power injection measurements and power flow measurements when $\lambda = 2$, $k = 2$, SNR = 30 dB and $\rho=0.1$ on IEEE 9-bus and 30-bus systems.}\label{fig12}
	\vspace{-5mm}
\end{figure}

\subsection{Assessment of vulnerability index (VuIx)}\label{Performance of VuIx for all the measurements}
Fig.~\ref{fig1} and Fig.~\ref{fig2} depict the mean and variance of the VuIx obtained from Algorithm~\ref{alg:vulnerability index} for all the measurements with SNR = 10 dB, $\lambda = 2$ and $\rho = 0.1$ on the IEEE 9-bus system when $k = 1$ and $k =2$, respectively. 
It is observed that power injection measurements yield higher priority vulnerability indices, which indicates that power injection measurements are more vulnerable to data integrity attacks. 
Most power injection measurements correspond to higher ranked vulnerability indices but there are instances of power flow measurements with a higher ranked VuIx than that of some power injection measurements. 
Interestingly, the power injection measurements with lower vulnerability indices correspond to the buses that are isolated in the system, that is, the buses with a lower number of connections. On the other hand, the power flow measurements with higher ranked vulnerability indices correspond to the branches with higher admittance. 
The VuIx for $k=0$ obtained in Corollary~\ref{coro_vulnerabilityordering_k0} is depicted for the purpose of  serving as a reference to assess the deviation when $k>0$. Interestingly, the VuIx of most measurements does not change significantly for different values of $k$, which suggests that the VuIx is insensitive to the state of the system.

Fig.~\ref{fig3} and Fig.~\ref{fig4} depict the mean and variance of the VuIx from Algorithm~\ref{alg:vulnerability index} for all the measurements with SNR = 30 dB, $\lambda = 2$ and $\rho = 0.1$ on the IEEE 9-bus system when $k = 1$ and $k = 2$, respectively. Interestingly, the mean of the VuIx for most of the measurements does not deviate significantly from the case when $k = 0$. Instead, most of the variances deviate significantly in comparison with the cases in Fig.~\ref{fig1} and Fig.~\ref{fig2} with SNR = 10 dB.
Fig.~\ref{fig5} and Fig.~\ref{fig6} depict the results on IEEE 30-bus systems with the same setting as in Fig.~\ref{fig1} and Fig.~\ref{fig2}, respectively. Fig.~\ref{fig7} and Fig.~\ref{fig8} depict the results on IEEE 30-bus systems with the same setting as in Fig.~\ref{fig3} and Fig.~\ref{fig4}, respectively. Surprisingly, the mean of the VuIx in larger systems coincides with that obtained for the case $k=0$, which shows that the VuIx is a robust security metric for large systems. Interestingly, the power injection measurements corresponding to the least connected buses decrease in the VuIx when SNR = 10 dB.

\subsection{Comparative vulnerability assessment of power flow and power injection measurements}

 
In Section~\ref{Performance of VuIx for all the measurements} we have established that power injection measurements and power flow measurements are qualitatively different in terms of the VuIx. 
To provide a quantitative description of this difference, Fig.~\ref{fig12} depicts the probability of a given VuIx $i\in\left\{1, 2, \ldots, m-|\Kc_a|\right\}$ being taken by a power injection measurement or a power flow measurement for the IEEE 9-bus and 30-bus systems when $\lambda = 2$, $k = 2$, SNR = 30 dB and $\rho=0.1$. Specifically, Fig.~\ref{fig12} depicts the probability of the following events:
\vspace{-1mm}
\begin{align*}
{\sf Flow}_i\!&\!: \textnormal{VuIx $i$ corresponds to a power flow measurement},\\
{\sf Inj}_i\!&\!: \textnormal{VuIx $i$ corresponds to a power injection measurement}.
\end{align*}

It is observed that in both systems, small VuIx are more likely to correspond to power injection measurements than to power flow measurements, that is, $\mathds{P}[{\sf Inj}_i]>\mathds{P}[{\sf Flow}_i]$ for small values of $i$. Conversely, it holds that $\mathds{P}[{\sf Inj}_i]<\mathds{P}[{\sf Flow}_i]$ for large values of $i$.
In fact, small VuIx corresponding to power injection measurements is with probability one, which shows that the most vulnerable measurements in the system are always power injection measurements. Conversely, the larger VuIx values corresponding to power flow measurements is with probability one, which indicates that the least vulnerable measurements are always power flow measurements. Interestingly, there is a clear demarcation for each system for which $\mathds{P}[{\sf Inj}_i]$ and $\mathds{P}[{\sf Flow}_i]$ change rapidly with the VuIx value, which suggests a phase transition type phenomenon for measurement vulnerability.

\section{Conclusion}\label{conclusion}
 
In this paper, we have designed a novel security metric referred to as vulnerability index (VuIx) that characterizes vulnerability of power system measurements to data integrity attacks from a fundamental perspective. We have achieved this by embedding information theoretic measures into the metric definition. The resulting VuIx framework evaluates the vulnerability of all the measurements in the systems and enables the operator to identify those that are more exposed to data integrity threats. We have tested the framework for IEEE test systems and concluded that power injection measurements are more vulnerable to data integrity attacks than power flow measurements. 

 \vspace{-2mm}
\bibliographystyle{IEEEtran}
\bibliography{SGC_22}

\begin{thebibliography}{10}
\providecommand{\url}[1]{#1}
\csname url@samestyle\endcsname
\providecommand{\newblock}{\relax}
\providecommand{\bibinfo}[2]{#2}
\providecommand{\BIBentrySTDinterwordspacing}{\spaceskip=0pt\relax}
\providecommand{\BIBentryALTinterwordstretchfactor}{4}
\providecommand{\BIBentryALTinterwordspacing}{\spaceskip=\fontdimen2\font plus
\BIBentryALTinterwordstretchfactor\fontdimen3\font minus
  \fontdimen4\font\relax}
\providecommand{\BIBforeignlanguage}[2]{{%
\expandafter\ifx\csname l@#1\endcsname\relax
\typeout{** WARNING: IEEEtran.bst: No hyphenation pattern has been}%
\typeout{** loaded for the language `#1'. Using the pattern for}%
\typeout{** the default language instead.}%
\else
\language=\csname l@#1\endcsname
\fi
#2}}
\providecommand{\BIBdecl}{\relax}
\BIBdecl

\bibitem{GJ_PSanalysis_1994}
J.~J. Grainger and W.~D. Stevenson, \emph{Power system analysis}.\hskip 1em
  plus 0.5em minus 0.4em\relax McGraw-Hill, 1994.

\bibitem{AA_PSstateestimation_04}
A.~Abur and A.~G. Exposito, \emph{Power system state estimation: Theory and
  implementation}.\hskip 1em plus 0.5em minus 0.4em\relax CRC press, Mar. 2004.

\bibitem{WZ_CN_13}
W.~Wang and Z.~Lu, ``Cyber security in the smart grid: Survey and challenges,''
  \emph{Computer networks}, vol.~57, no.~5, pp. 1344--1371, Jan. 2013.

\bibitem{JA_PearsonEducation_07}
A.~Jaquith, \emph{Security metrics: replacing fear, uncertainty, and
  doubt}.\hskip 1em plus 0.5em minus 0.4em\relax Pearson Education, 2007.

\bibitem{M_CVSS_11}
\BIBentryALTinterwordspacing
M.~Schiffman, ``Common \ce{V}ulnerability \ce{S}coring \ce{S}ystem
  (\ce{CVSS}),'' 2011. [Online]. Available:
  \url{\url{http://www.first.org/cvss/cvss-guide.html}}
\BIBentrySTDinterwordspacing

\bibitem{NISTIR_securityguideline_10}
\BIBentryALTinterwordspacing
V.~Y. Pallitteri and T.~L. Brewer, \emph{Guidelines for Smart Grid
  Cybersecurity}.\hskip 1em plus 0.5em minus 0.4em\relax NIST
  Interagency/Internal Report (NISTIR), National Institute of Standards and
  Technology, 2014. [Online]. Available:
  \url{\url{https://doi.org/10.6028/NIST.IR.7628r1}}
\BIBentrySTDinterwordspacing

\bibitem{PM_ACM_16}
M.~Pendleton, R.~Garcia-Lebron, J.~H. Cho, and S.~Xu, ``A survey on systems
  security metrics,'' \emph{ACM Computing Surveys}, vol.~49, no.~4, pp. 1--35,
  Dec. 2017.

\bibitem{VV_TSG_19}
V.~Venkataramanan, A.~Hahn, and A.~Srivastava, ``\ce{CP-SAM}: Cyber-physical
  security assessment metric for monitoring microgrid resiliency,'' \emph{IEEE
  Trans. on Smart Grid}, vol.~11, no.~2, pp. 1055--1065, Mar. 2022.

\bibitem{LY_TISSEC_11}
Y.~Liu, P.~Ning, and M.~K. Reiter, ``False data injection attacks against state
  estimation in electric power grids,'' \emph{ACM Trans. Info. Syst. Sec},
  vol.~14, no.~1, pp. 1--33, May 2011.

\bibitem{CKKPT_SPM_12}
S.~Cui, Z.~Han, S.~Kar, T.~T. Kim, H.~V. Poor, and A.~Tajer, ``Coordinated
  data-injection attack and detection in the smart grid: A detailed look at
  enriching detection solutions,'' \emph{IEEE Signal Process. Mag}, vol.~29,
  no.~5, pp. 106--115, Aug. 2012.

\bibitem{MO_JSAC_13}
M.~Ozay, I.~Esnaola, F.~T.~Y. Vural, S.~R. Kulkarni, and H.~V. Poor, ``Sparse
  attack construction and state estimation in the smart grid: Centralized and
  distributed models,'' \emph{IEEE J. Sel. Areas Commun.}, vol.~31, no.~7, pp.
  1306--1318, Jul. 2013.

\bibitem{IE_TSG_16}
I.~Esnaola, S.~M. Perlaza, H.~V. Poor, and O.~Kosut, ``Maximum distortion
  attacks in electricity grids,'' \emph{IEEE Trans. Smart Grid}, vol.~7, no.~4,
  pp. 2007--2015, Jul. 2016.

\bibitem{SE_TSG_19}
K.~Sun, I.~Esnaola, S.~M. Perlaza, and H.~V. Poor, ``Stealth attacks on the
  smart grid,'' \emph{IEEE Trans. Smart Grid}, vol.~11, no.~2, pp. 1276--1285,
  Aug. 2019.

\bibitem{YE_SGC_20}
X.~Ye, I.~Esnaola, S.~M. Perlaza, and R.~F. Harrison, ``Information theoretic
  data injection attacks with sparsity constraints,'' in \emph{Proc. IEEE Int.
  Conf. on Smart Grid Comm.}, Tempe, AZ, USA, Nov. 2020.

\bibitem{GE_TSG_18}
C.~Genes, I.~Esnaola, S.~M. Perlaza, L.~F. Ochoa, and D.~Coca, ``Robust
  recovery of missing data in electricity distribution systems,'' \emph{IEEE
  Trans. Smart Grid}, vol.~10, no.~4, pp. 4057--4067, Jun. 2018.

\bibitem{SI_TIT_13}
I.~Shomorony and A.~S. Avestimehr, ``Worst-case additive noise in wireless
  networks,'' \emph{IEEE Trans. Inf. Theory}, vol.~59, no.~6, pp. 3833--3847,
  Jun. 2013.

\bibitem{book_EIT}
T.~M. Cover and J.~A. Thomas, \emph{Elements of information theory}.\hskip 1em
  plus 0.5em minus 0.4em\relax John Wiley \& Sons, 1999.

\bibitem{UoW_ITC_99}
\BIBentryALTinterwordspacing
U.~of~Washington, ``Power systems test case archive,'' 1999. [Online].
  Available:
  \url{\url{https://sentinel.esa.int/web/sentinel/user-guides/sentinel-2-msi/resolutions/radiometric}}
\BIBentrySTDinterwordspacing

\bibitem{matpower}
R.~D. Zimmerman, C.~E. Murillo-S{\'a}nchez, and R.~J. Thomas, ``Matpower:
  Steady-state operations, planning, and analysis tools for power systems
  research and education,'' \emph{IEEE Trans. Power Syst}, vol.~26, no.~1, pp.
  12--19, Feb. 2010.

\end{thebibliography}

\end{document}